# Ideal Free Distribution in Agents with Evolved Neural Architectures

Virgil Griffith, Larry S. Yaeger

Indiana University School of Informatics and
Department of Cognitive Science
griffitd@indiana.edu

**Abstract**

We investigate the matching of agents to resources in a computational ecology configured to present heterogeneous resource patches to evolving, neurally controlled agents. We repeatedly find a nearly optimal, ideal free distribution (IFD) of agents to resources. Deviations from IFD are shown to be consistent with models of human foraging behaviors, and possibly driven by spatial constraints and maximum foraging rates. The lack of any model parameters addressing agent foraging or clustering behaviors and the biological verisimilitude of our agent control systems differentiates these results from simpler models and suggests the possibility of exploring the underlying mechanisms by which optimal foraging emerges.

## Introduction

A key question in contemporary ecology is, How should organisms distribute themselves among patches of differing quality? This question was originally addressed in the Ideal Free Distribution (IFD) model of Fretwell and Lucas (1970, Fretwell 1972). The IFD model predicts that in an environment in which food resources are distributed heterogeneously amongst multiple patches, the proportion of organisms in each patch will be identical to the proportion of food in each patch.

Deviations from IFD are widely documented (Parker and Sutherland 1986, Kennedy and Gray 1993, Tregenza 1995, Roberts and Goldstone 2005), and arise from various violations of IFD assumptions. Nonetheless, IFD continues to be a valuable theoretical and modeling tool, and has been extended along almost every dimension previously considered a violation of its assumptions: unequal competitors (Rosenzweig 1986, Korona 1989, Sutherland and Parker 1992); resource dynamics (Schwinning and Rosenzweig 1990, Lessells 1995); assessment rules and perceptual constraints (Abrahams 1986, Bernstein et al. 1988); travel costs (Tyler and Gilliam 1995); patch memory and selection (Milinski 1994); spatial constraints (Stephens and Stevens 2000, Shepherd and Litvak 2004).

Empirical studies show at least approximate agreement between IFD and the behavior of many biological organisms, from dungflies (Parker 1974, 1978) to humans (expressed as a "matching law"; Herrnstein 1961). The review by Tregenza (1995) lists 48 empirical studies of IFD from 1970 to 1995.

In this paper we investigate habitat matching in the Polyworld (Yaeger 1994) computational ecology, in which the agents' behaviors are driven entirely by artificial neural networks, with evolved neural architectures. We examine these agents' compliance with and deviation from IFD, and the conditions under which these ranges of behavior occur.

## Tools and Techniques

**Polyworld Simulator**

Polyworld is a computational ecology where simulated haploid organisms compete for survival. The environment is a continuous (un-gridded) and bounded plane populated by agents and food. Polyworld agents have a suite of seven primitive behaviors: move, turn, eat, mate, attack, light, and focus. Their behaviors are exclusively controlled through the continuous activations of artificial neural networks (ANNs). The ANNs consist of summing and squashing neurons (roughly 50 to 200 in this series of runs), that employ Hebbian learning at the synapses.

The architecture of the ANNs are probabilistically derived from the organisms' virtual genes. A given genome produces a small class of neural anatomies. The full range of genetic encodings produces a very large class of neural anatomies. Evolution of neural anatomies is the main focus of Polyworld, and there is evidence that these networks exhibit an evolutionary trend towards greater complexity of neural structure and function (Yaeger and Sporns 2006).

The primary input to the agents' ANNs, and their only real "sense" mechanism is vision, via a 1D strip of pixels taken from a rendered image of their environment. The 3D environment is rendered from each agent's point of view and the resulting pixel map is fed as input to the ANN as though it were light falling on a retina. The agents' range of vision is effectively infinite, except where occluded by foreground objects.

In Polyworld, all actions expend energy, including neural activity. Agents also lose energy at a low rate, independent of activity. Thus, in order to persist, agents must find food to eat. They must also find mating partners with which to reproduce.

Polyworld populations may be self-sustaining, via their mating behaviors, or not. When they *are* self-sustaining, there is no fitness function; evolution is "guided" purely by natural selection. Whenever a population is *not* self-sustaining, newly created agents are introduced, as necessary, to maintain a minimum population. This continual re-seeding is accomplished by mating pairs of agents from an N-best list, where "best" is determined by an *ad hoc* heuristic fitness function.

To limit computational demands, a maximum population size of 300 is imposed. This limit can have profound evolutionary consequences, by preventing otherwise viable agents from producing offspring, and thus reducing or eliminating evolutionary change. To mitigate this effect, we maintain a list of the *least* fit agents as determined by the heuristic fitness function, and, while at maximum population, when two organisms attempt to mate, the agent with the lowest fitness is killed, allowing the new offspring to be born. This *replace least fit* technique is common in genetic algorithms, and partially overcomes the slowdown in evolutionary change resulting from the population limit.

Additional details of the simulation engine can be found in (Yaeger 1994), or by consulting the source code at http://sourceforge.net/projects/polyworld.

**Experimental Design**

In all simulations, we create a world of equal width and depth (100x100 units). The world has two resource patches that are maximally distant from each other, one at each of the near and far ends of the world. Each resource patch consists of a band running the entire width of the world, in which food is distributed randomly. Food bands cover exactly half of the world (50 units), with the remaining 50 units forming a food-free region between the food bands.

Food is distributed between the two bands according to a specifiable, per-band *food fraction*. Except as noted, food is distributed *probabilistically*; i.e., when a piece of food is added to the world, a band's food fraction is the probability that the new piece of food will be placed in that band. Initially these food fractions correspond precisely to the amount of food in each band. Over time, however, over-foraging in one band and under-foraging in the other, may cause the actual proportion of food in the bands to vary.

Figure 1 shows three views of the simulation environment, at early, middle, and late stages of evolution. (Images have false color and altered perspective to aid clarity.) The white blocks are food; the gray trapezoids are agents. Upon initialization (1A), the food is mostly on the distant end of the world and the agents are uniformly distributed. At a middle stage of evolution (1B), there are more agents, mostly occupying and over-foraging the distant food patch, while the close food patch has been under-foraged and currently has more food than its assigned food fraction would indicate. At a late stage of evolution (1C), agent and food distributions are approximately equal to the assigned food fractions.

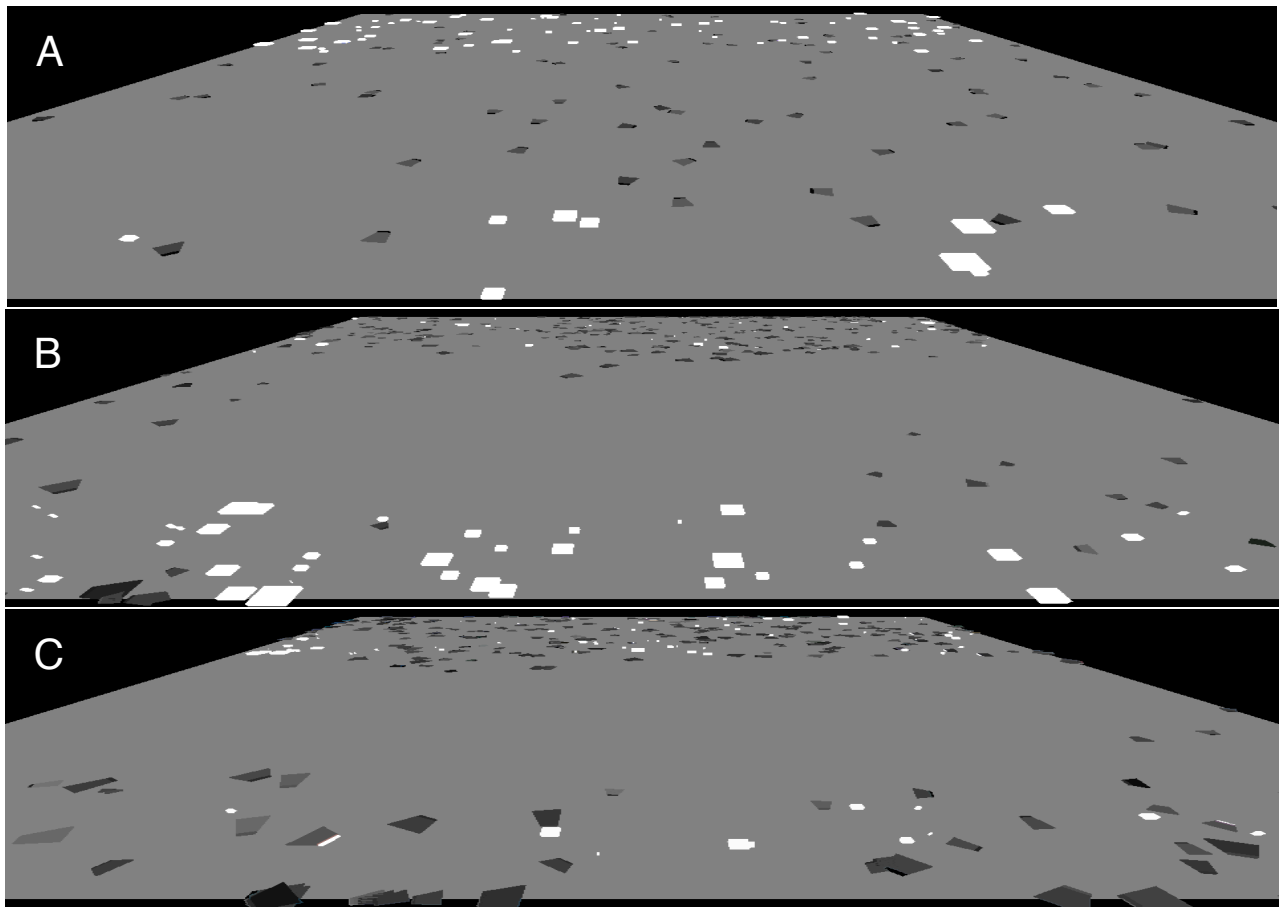

Figure 1: (A) Early, (B) middle, and (C) late views of a simulation

We also ran a few experiments in which food is distributed *deterministically* or *rigidly*. I.e., when a piece of food is added to the world, we guarantee that the fraction of food in each band remains fixed. These experiments are clearly identified and discussed separately.

We begin these simulations with 90 clones of a "progenitor" ancestor, in order to promote evolutionary convergence at comparable rates across all simulations. This, in effect, constricts the massive phenotypic/genotypic state space that evolution is likely to explore. The progenitor genome was designed so as to elicit a limited set of nominal, "reasonable" agent behaviors—move towards green (food), turn away from and attack red (attacking agents), and attempt to mate with blue (mating agents). However, the progenitor ancestors do not form a self-sustaining population without being subjected to variation and selection. Starting with random genomes produces essentially identical results, but on highly variable time scales.

We run all simulations for 10,000 time steps (about 30 times the lifespan of a typical agent, 200 times the minimum time between matings). Agents capable of sustaining a maximum population, distributed amongst the two food bands, emerge in less than 5,000 time steps. We continue recording to 10,000 time steps to assess the stability of these agent distributions.

Every 10 time steps we collect numerous statistics about the environment. For this exercise, the most important of these are: the number of agents in each food band, the amount of food in each band, the number of agents *near* but technically not in either band, the number of agents not in *or near* either band, and the total population size. We then vary the size and/or food fraction of each band and observe how the evolved agents distribute themselves between the two resource patches.

Evolution is free to generate neural architectures that elicit behaviors geared to perception of food, other agents, potential mates, etc., but no further external intervention is allowed, and there are no model parameters directly controlling agent foraging or flocking behaviors.

## Results

When assessing a population's match to IFD, it is common to consider only the relative number of agents in each patch, ignoring agents that are not in any patch. Accordingly, the majority of our reported agent distributions are *normalized*, based on the sum of agents in any patch, rather than the total number of agents. However, since we are able to account for the whereabouts of every member of our population (unlike studies of natural habitats), we also examine a sample of raw, un-normalized data to better understand the actual distribution of agents.

With spatially instantiated food patches and mobile agents, there is also an issue of how to account for agents that spend much of their time just outside a resource patch but frequently skirt inside. Some food pieces may also be wide enough to extend slightly beyond the borders of the designated food bands, further complicating the issue. Since the normalized data ignores out-of-band agents, we ignore these edge-skirters in the majority of our analysis, but will look briefly at the magnitude of this effect when we examine the raw, un-normalized agent distributions.

Figure 2 shows the results of a series of runs with food bands of equal size but with unequal food fractions. This produces two bands with the same area but with different food densities. All runs show agent distributions that are very close to an IFD, including the limiting conditions of 100% / 0% and 50% / 50%.

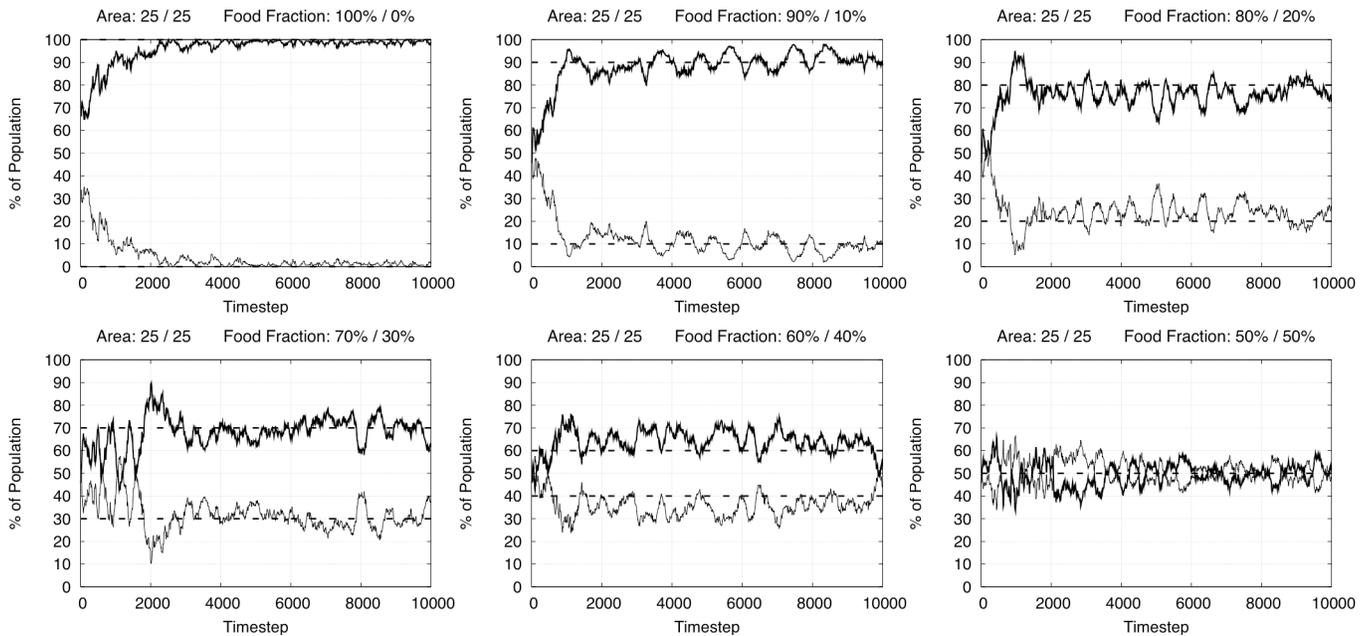

Figure 2: Normalized agent distributions in two food bands with equal areas and unequal densities. The darker line is the higher density band. IFD is shown in dashed lines.

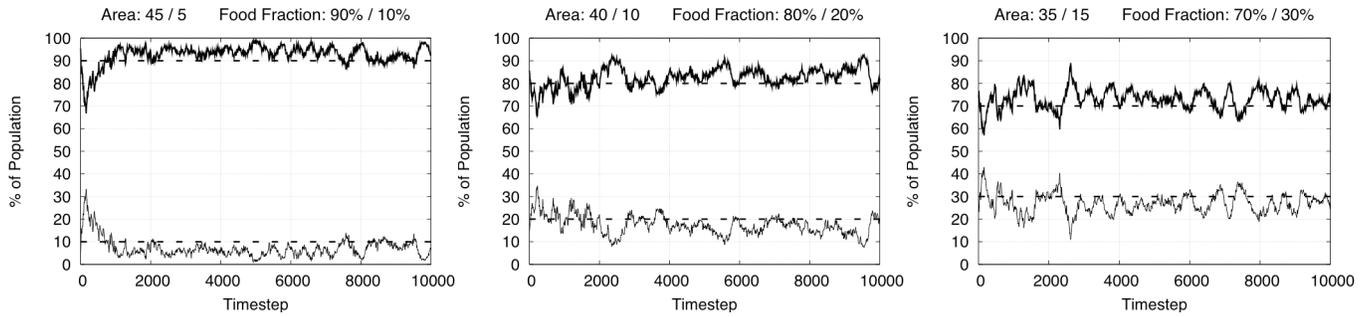

Figure 3: Normalized agent distributions in two food bands with equal densities and unequal areas. The darker line is the larger area band. IFD is shown in dashed lines.

Figure 3 shows a series of runs with bands of equal food density, but unequal area. Again all runs result in evolved agent distributions that approximately match the distribution of resources, though there may be a suggestion of consistent, albeit slight overmatching (more agents in the large patch than IFD would predict). Results from distributions not shown are consistent with those that are.

Figure 4 shows raw, un-normalized data from a single run with equal areas and unequal food densities (the same run that produced the normalized results for these conditions in Figure 2). In addition to agent populations of the two food bands, an additional, lighter curve shows the percentage of agents not in any band. Though accounting for agents that are not in any band affects the distributions, the resulting populations still clearly exhibit near-optimal foraging. These results are typical.

Figure 4 also informs the issue of edge-skirters. In addition to the strict in-band agent counts, that serve as lower bounds, we measured agent counts that include a 10 unit "border zone" adjacent to the food bands. This produces an upper bound on the number of agents that might reasonably be considered part of a food band. The thickness of the in-band agent-distribution curves thus conveys a range of possible estimates of band membership. (The bottom, out-of-band curve is thicker because it combines the effects of both border zones.) As with normalization, though details of the distributions are affected by different methods of accounting for edge-skirters, the resulting distributions remain nearly optimal.

The results for both the equal area/unequal density and equal density/unequal area cases, normalized and un-normalized, are summarized in Table 1.

**Over- and Under-Foraging**

Due to variable foraging efficiency, the percentage of food in the two bands can vary substantially from the specified food fractions. Figure 5 looks at the actual percentages of agents and food in the high-density band of an equal area/unequal density experiment.

Note that every peak in the number of agents is immediately followed by a dip in food density, which leads to a dip in the number of agents, whereupon food density recovers, and so on. The result is an irregular, but otherwise classic Lotka-Volterra predator/prey cycle between foragers and food. Clearly the patch depletion that results from over-foraging and the patch abundance that results from under-foraging is producing strong shifts in the actual availability of food in the two bands, which turns out to be crucial to the evolution of an IFD in our simulated environment.

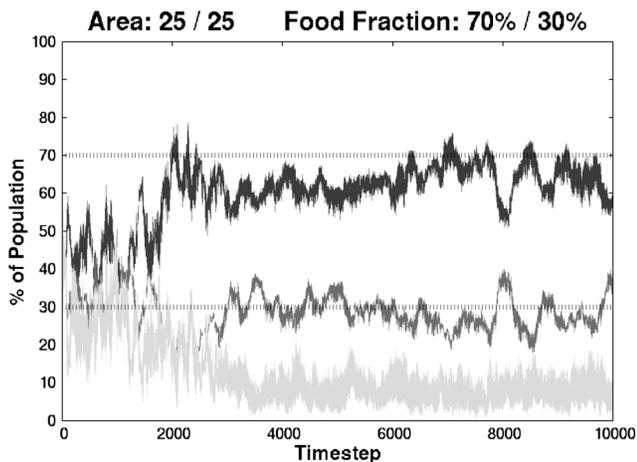

Figure 4: Raw (un-normalized) agent distributions in two food bands with equal areas and unequal densities. Bottom curve shows agents not in any band.

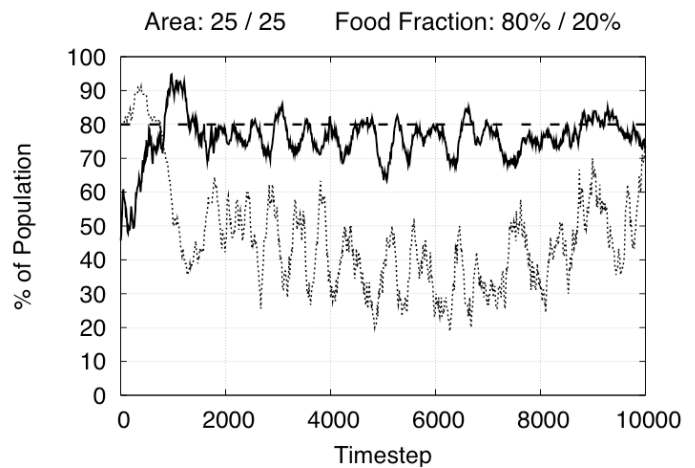

Figure 5: Normalized agent and food occupancy of the high-density band.

| Band sizes | 25/25 | 30/20 | 35/15 | 40/10 | 45/5 | 50/0 |
| --- | --- | --- | --- | --- | --- | --- |
| Food fractions (as %) | 50/50 | 60/40 | 70/30 | 80/20 | 90/10 | 100/0 |
| Agents | 43/43/14 | 53/33/14 | 63/23/14 | 73/14/13 | 80/6/14 | 92/0/8 |
| Agents+10 | 48/48/4 | 58/37/5 | 68/27/5 | 78/18/4 | 86/9/5 | 99/0/1 |
| Agents (normalized) | 50/50 | 62/38 | 73/27 | 84/16 | 94/6 | 100/0 |
| Band sizes | 25/25 | 25/25 | 25/25 | 25/25 | 25/25 | 25/25 |
| Food fractions (as %) | 50/50 | 60/40 | 70/30 | 80/20 | 90/10 | 100/0 |
| Agents | 43/43/14 | 56/31/13 | 61/26/13 | 66/21/13 | 80/8/12 | 91/1/8 |
| Agents+10 | 48/48/4 | 62/34/4 | 66/30/4 | 73/23/4 | 87/10/3 | 97/2/1 |
| Agents (normalized) | 50/50 | 65/35 | 70/30 | 76/24 | 90/10 | 99/1 |

Table 1: Agent distributions for various combinations of food band areas and food fractions. Top: Equal areas, with unequal densities. Bottom: Unequal areas, with equal densities. Table shows agent percentages in high-density (or area) band / low-density (or area) band/ not in any band (for un-normalized data only).

### Rigid Food Distribution

Having observed this substantial impact of foraging on food distribution, we decided to implement a more rigid, deterministic food distribution algorithm. With rigid food distribution, food is always replenished in the band where it was eaten, thus guaranteeing that the specified and actual food fractions are the same and that they remain constant. Figure 6 shows the normalized agent distributions for food bands with equal area and unequal densities using this deterministic food distribution scheme. All plots show extreme overmatching. This phenomenon is rarely seen in the literature, however, in a private communication Roberts and Goldstone observed that their model of resource matching also showed extreme over-matching when using a deterministic update rule. This is in contrast to the predominantly under-matching results they obtained with the probabilistic food distribution scheme employed in their published results (Roberts and Goldstone 2005).

### Undermatching and Overmatching

Though the bulk of Roberts and Goldstone's reported results involve resource pools with a Gaussian distribution of food, they discuss three tests in which food is distributed with uniform variance in the food pools, more like the experiments being reported here. Figure 7 shows the agent distributions in three Polyworld runs corresponding to the conditions tested with Roberts and Goldstone's EPICURE model. The first, left-most plot in both sets of results is for food pools (or bands) that are of the same area, but differ in food density; both models show IFD resource matching or very slight undermatching. The middle plots are for two food pools with differing areas, but equal food density; both models show a slight overmatching, though ours is less pronounced. The final, right-most plots are for a pair of pools that differ in both area and food density. Here, a much smaller pool has a much higher food density. Under this somewhat unusual condition, both models show unusually strong undermatching, though again EPICURE's deviation from IFD is more pronounced.

### Discussion

The evolved, neurally controlled agents of Polyworld clearly exhibit a nearly Ideal Free Distribution of agents to resources, when the food is distributed probabilistically. We observe near-IFD broadly and consistently, despite the absence of any explicit rules guiding the agents' behaviors.

Without patch depletion effects—without over-foraging and under-foraging—our agents do not exhibit an IFD, instead showing essentially pure overmatching, with all agents coming to reside in the higher-density patch. Intriguingly, similar overmatching is observed in Roberts and Goldstone's rule-based EPICURE model of resource matching in humans, once they switch to the same kind of rigid food distribution policy. We believe this phenomenon follows from the higher probability of finding a piece of food in the higher density patch. Certainly the probability of encountering a piece of food per time step is higher in the high-density patch initially, and with perfect and complete food replacement, that probability remains higher, despite increasing agent density. Since food is replaced only once per time step in Polyworld, perhaps if agent density increased to the point that a substantial fraction of all food in the patch was eaten in a single time step we would see something other than extreme overmatching with rigid food distribution.

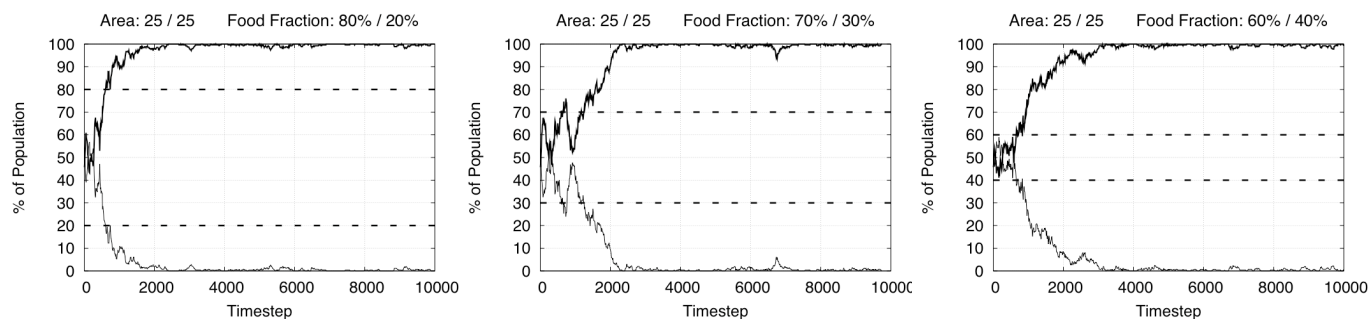

Figure 6: Normalized agent distributions in two food bands with equal areas and unequal densities, using "rigid" food distribution. The darker line is the higher density band. IFD is shown in dashed lines.

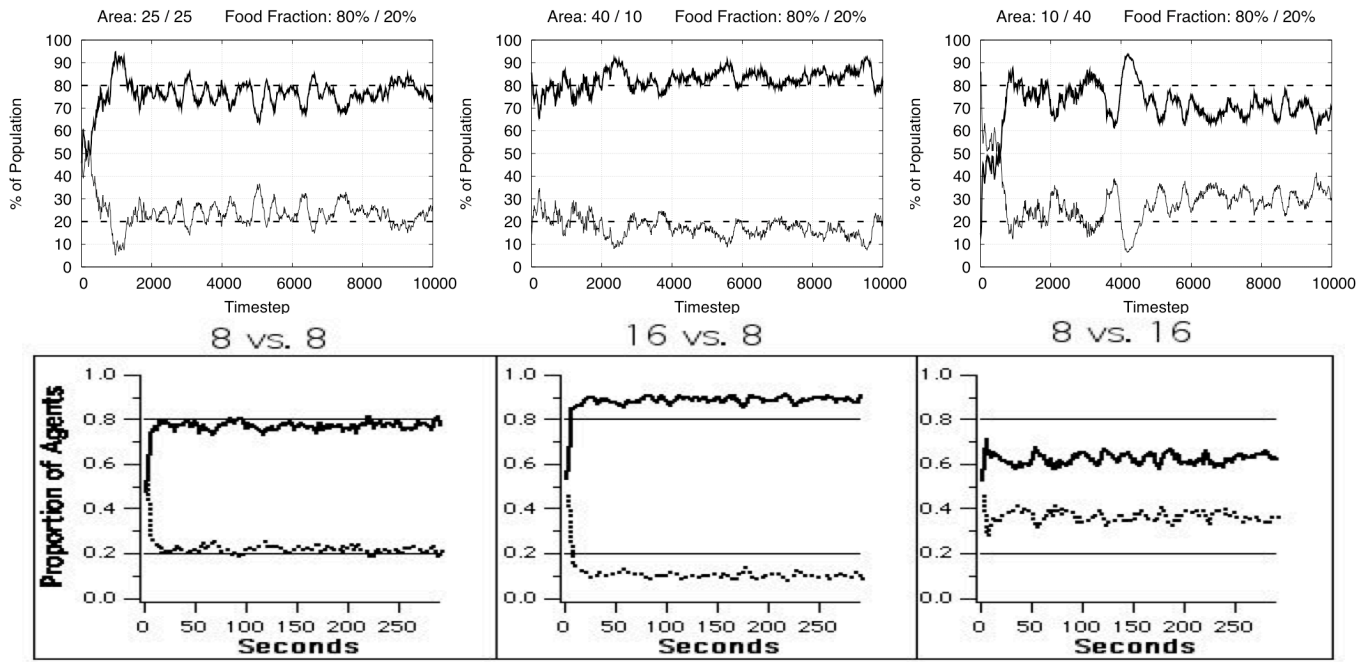

Figure 7: Normalized agent distributions for three different food area and density combinations. The darker line is the higher density band (larger area if densities are the same). Top row are Polyworld's evolved neural agents. Bottom row are Roberts and Goldstone's EPICURE model agents using comparable area and density ratios (Figure 5 from Roberts and Goldstone 2005, using visible food and visible agents, with permission).

Deviations from IFD for the probabilistic food update model are slight, but again follow the same trends as the EPICURE model of human resource matching. Equal area, but unequal density patches match IFD fairly closely. Unequal areas, but equal density patches produce a slight overmatching. Extreme disparity in densities, when combined with a reverse disparity in areas, produces a substantial undermatching to resources. Even though agents in Polyworld can overlap each other completely, there may be a sufficient density of agents in this case that the agents' vision is obscured, thus diminishing their ability to locate pieces of food in the high-density patch, and making the more isolated pieces of food, clearly visible in the low-density patch, more attractive. It is for very different underlying reasons, but Roberts and Goldstone speculate that their model's undermatching in this case may also be driven by spatial constraints and agent density.

We have repeatedly referred to Roberts and Goldstone's work here because we are frankly amazed at the strong consistency and agreement between two such radically different agent and resource models. Yet perhaps we shouldn't be. EPICURE's design provides sensible rules about foraging behaviors. Polyworld evolves sensible foraging behaviors.

## Future Directions

There are a variety of questions raised and left unanswered by this work. From a theoretical perspective, the probability of encountering a unit of food per unit of time, and the corresponding predictions of agent distributions, should be relatable to direct measurements of either food consumed per agent per time step or pieces of food encountered per agent per time step. Both measurements should be straightforward to acquire, and making them should help us understand the directions and magnitudes of the observed deviations from IFD, and provide a formal bridge to the emergence of such distributions and optimal foraging behaviors in biological organisms.

More empirically, it should be straightforward to investigate the effects of travel time on agent distributions and their deviations from IFD. If properly framed, it would be almost as simple to investigate the effect of predators. And with a modest change to the food requirements of the agents it would be feasible to investigate the effects on IFD of minimum resource requirements in a varied diet.

## Acknowledgements

Thanks to Rob Goldstone, Michael Roberts, and Peter Todd for valuable and enlightening conversations.